\input{aipcheck}

\documentclass[
    ,final            % use final for the camera ready runs
  ]
  {aipproc}

\layoutstyle{6x9}

\usepackage{amsmath}

\def\join{\mathop\vee}
\def\meet{\mathop\wedge}
\def\bor{\mathop{\mathord{\lor}\!\!\!\raise4pt\hbox{$\scriptscriptstyle 2$}\,}}
\def\band{\mathop{\mathord{\land}\!\!\!\lower2pt\hbox{$\scriptscriptstyle 2$}\,}}

\begin{document}

\title{Information Physics:\\The New Frontier}


\author{Kevin H. Knuth}{address={University at Albany (SUNY), NY, USA\\ Email: kknuth@albany.edu}}

\begin{abstract}
At this point in time, two major areas of physics, statistical mechanics and quantum mechanics, rest on the foundations of probability and entropy.  The last century saw several significant fundamental advances in our understanding of the process of inference, which make it clear that these are inferential theories.  That is, rather than being a description of the behavior of the universe, these theories describe how observers can make optimal predictions about the universe.  In such a picture, information plays a critical role.  What is more is that little clues, such as the fact that black holes have entropy, continue to suggest that information is fundamental to physics in general.

In the last decade, our fundamental understanding of probability theory has led to a Bayesian revolution.  In addition, we have come to recognize that the foundations go far deeper and that Cox's approach of generalizing a Boolean algebra to a probability calculus is the first specific example of the more fundamental idea of assigning valuations to partially-ordered sets.  By considering this as a natural way to introduce quantification to the more fundamental notion of ordering, one obtains an entirely new way of deriving physical laws.  I will introduce this new way of thinking by demonstrating how one can quantify partially-ordered sets and, in the process, derive physical laws.  The implication is that physical law does not reflect the order in the universe, instead it is derived from the order imposed by our description of the universe.  Information physics, which is based on understanding the ways in which we both \emph{quantify} and \emph{process} information about the world around us, is a fundamentally new approach to science.
\end{abstract}

\maketitle

%%%%%%%%%%%%%%%%%%%%%%%%%%%%%%%%%%%%%%%%%%%%
%% MAINMATTER
%%%%%%%%%%%%%%%%%%%%%%%%%%%%%%%%%%%%%%%%%%%%

\begin{quote}
``Measure what is measurable, and make measurable what is not so.''\\
Galileo Galilei (1564-1642)
\end{quote}

\section{Introduction}
In the last century, there were three individuals whose ideas revolutionized the way we view information and probability.  The first of these individuals was Claude Shannon who, while in graduate school, realized that Boolean algebra could be used to simplify telephone networks.  This insight paved the way for digital computers, which clearly have revolutionized all aspects of human society.  However, it also led to a more subtle revolution based on Shannon's quantification of information transmitted by a communication channel.  Shannon's information took the curious form of entropy \cite{Shannon&Weaver}, which at the time was believed to be a physical property of a thermodynamic system.

Around the same time, a physicist, Richard Threlkeld Cox, published a paper where he obtained probability theory as a unique quantification of degrees of plausibility deriving from a generalization of Boolean algebra \cite{Cox:1946}.  To this day, Cox's results are not fully appreciated by the scientific community.  His approach forms a foundation for probability theory that stands alongside of the measure-theoretic foundation provided by Kolmogorov.  While Kolmogorov's approach is founded in traditional mathematical rigor, Cox's approach relies on a purpose-driven generalization, which is perhaps more satisfying to physicists, but less so to mathematicians.  However, the motivation behind the specific generalization that Cox proposes gives \emph{meaning} to the concept of probability, which is something that Kolmogorov's approach lacks.  As Bayesians, we often view probabilities as degrees of plausibility, or degrees of belief, and many of us have come to find Cox's views quite natural.

Edwin T. Jaynes discovered Shannon's paper in the Princeton library, and as he says, he disappeared for about a week \cite{Jaynes:1956}.  Upon re-emerging, he declared to anyone who would listen that this was the greatest piece of work since the discovery of the Dirac equation.  Jaynes writes,
\begin{quote}
It's almost impossible to describe the psychological effect of seeing our old familiar expression for entropy derived in a completely new way, and then applied with great success to problems of engineering which apparently have no relation to thermodynamics.  But all of the inequalities, which are usually associated with the second law of thermodynamics, turn out to be statements of the greatest practical usefulness in engineering problems.  It seemed to me that there must be something pretty important that we could learn from this situation. \cite[p.~3]{Jaynes:1956}
\end{quote}

Many of the early attempts to employ information theory in physics were based on making analogies between the communication theory and statistical mechanics.  Jaynes realized that the connection was not in the form of a simple analogy, but was something far more subtle.  He writes
\begin{quote}
the essential content of both statistical mechanics and communication theory, of course, does not lie in the equations; it lies in the ideas that lead to those equations. \cite[p.~4]{Jaynes:1956}
\end{quote}
Jaynes continues by writing
\begin{quote}
the job as I saw it was not to try to invent any fancy new mathematics.  That would presumably come later if we were successful.  The job was to find the \underline{viewpoint} from which we could see that the reasoning behind communication theory and statistical mechanics was really the same. \cite[p.~5]{Jaynes:1956}
\end{quote}
This critical insight will be relevant again when we look at extending these ideas to quantum mechanics and beyond.

Jaynes was also aware of Cox's work in 1956 when he gave his lectures on Probability Theory in Science and Engineering.  Jaynes appreciated Cox's approach as it made clear that probability quantified a state of belief about a physical system rather than the state of the physical system itself.  He recognized that the latter viewpoint, led to potential misconceptions when probability theory was applied in physics.  While he was clearly convinced of the interpretation of probability as a degree of plausibility, he, like many of us, was not satisfied with Cox's derivation of the product rule.  Jaynes writes
\begin{quote}
I might say that I am not entirely satisfied with the argument that we went through to get this; not because I think its wrong, but because I think it is too long.  The final result we get is so simple that there must be a simpler way of deriving it; but I haven't found it. \cite[p.~35]{Jaynes:1956}
\end{quote}
A year after his lectures on the topic, Jaynes published his paper revealing the ideas behind both communication theory and statistical mechanics, which results in the principle of maximum entropy \cite[pp.~110---151]{Jaynes:1956}, \cite{Jaynes:InfoTheory}.  Since the entropy quantifies the degree of uncertainty in a probability distribution, assigning a probability that maximizes the entropy subject to a set of constraints amounts to using the information provided by the known constraints, while being careful not to inadvertently assume too much.  Jaynes' maximum entropy principle provided the justification that Gibbs so carefully avoided in his works on statistical mechanics to ensure acceptance.

With the benefit of the insights provided by these three individuals, we have come to view probability, entropy and information in a new light.  Probability and entropy describe states of knowledge about systems---not the systems themselves.  What is more, we now realize that information acts a constraint on our beliefs.  Free from the previous confusion surrounding probability, entropy and information, and the misconceptions that ensue, we can take these new \emph{ideas} and re-examine the laws of physics.  Several of us from this community have been doing just that.  In addition to a more clear understanding of statistical mechanics we have seen the principle of maximum entropy used to derive properties of systems ranging from the physics of foam \cite{Rivier-etal:foams} to the physics of planetary atmospheres \cite{Lorenz-etal:maxent-atmospheres}.  More profound perhaps is Ariel Caticha's investigation of entropic dynamics \cite{Caticha:Entropic_Dynamics} where he is working to utilize maximum entropy to derive the dynamical behavior of systems ranging from Newtonian mechanics \cite{Caticha:ED-and-Newtonian} to quantum mechanics \cite{Caticha:ED-and-QM}.

Inspired by Cox, I have been working to understand how to derive calculi from algebras in general by selecting consistent quantification schemes for partially-ordered sets and lattices.  At one level, this more fundamental understanding has resulted in a much simpler derivation of the product rule that might have been more to Jaynes' liking.  However, at a deeper level, we now understand how constraints imposed by ordering relations can result in the derivation of physical laws.  This recently has been demonstrated with a novel derivation of the complex arithmetic in Feynman's path integral approach to quantum mechanics \cite{GKS:me09, GKS:PRA} as well as a derivation of special relativity from a partial order on a set of events \cite{Knuth+Bahreyni:SR}.  Each of these examples is related to information in a different way.  In some examples the connection to information is direct as we consider a partial order on states of knowledge themselves.  However, we have also employed these ideas by considering the partial order that arises from the way that events can be informed about one another or the partial order that arises from composing sequences of measurements aimed at gaining information.

In this tutorial, which is still very much a work in progress, I will introduce this new way of thinking by explaining how one can derive physical laws by quantifying partially-ordered sets.  The implication is that physical law does not reflect the order in the universe, instead it is derived from the order imposed by our description of the universe.  This occurs both through the acts of \emph{quantification} of information (which I will discuss here) and \emph{processing} of information, which is related to the use of entropy and probability.   We have now demonstrated these ideas by deriving a surprising amount of old physics.  New physics now awaits as we enter this new frontier of Information Physics.

\subsection{Order Theory, Posets, Lattices and Algebras}
While group theory has become an essential tool for theoretical physics, order theory remains entirely overlooked.  At the most fundamental level, group theory is concerned with equivalence relations among partitioned sets, whereas order theory is concerned with ordering relations among ordered sets.  In this sense these two theories stand side-by-side and both can place extremely strong constraints on physical theories.  I will use these theories in concert with one another.  First, I will rely on ordering relations to obtain algebraic operations that have specific symmetry properties.  I will then use these symmetries to place strong constraints on any quantified description.  The resulting constraints correspond to the physical laws.

\begin{figure}[b]
  \label{fig:posets}
  \includegraphics[height=.15\textheight]{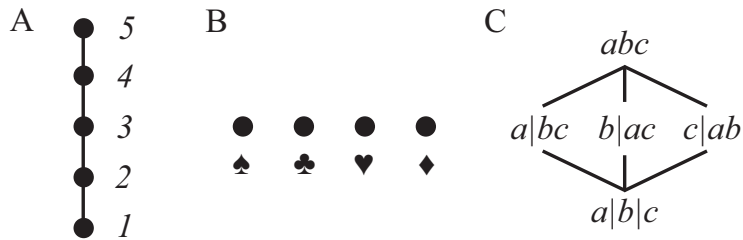}
  \caption{Three basic examples of posets. (A) The integers ordered by the usual $\leq$ form a \emph{chain}.  The element $2$ is drawn above $1$ since $1 \leq 2$, and they are connected by a line because $2$ covers $1$ in the sense that there is no integer $x$ between $2$ and $1$ such that $1 \leq x \leq 2$.  (B) The four card suits are incomparable under a wide variety of card game rules and we draw them side-by-side to express this.  This configuration is called an \emph{antichain}.  (C) The set of partitions of three elements $a$, $b$ and $c$ ordered by partition containment forms a more complex poset that exhibits both chain and antichain behavior.  One chain consists of the elements $a | b | c$, $a | bc$, and $abc$ since each successive partition contains the previous.  The elements $a | bc$, $b | ac$, and $c | ab$ form an antichain because not one of these three partitions contains another.}
\end{figure}

I begin by introducing the concept of a binary ordering relation and a partially-ordered set.  Two elements of a set are ordered by comparing them according to a binary ordering relation, generically denoted $\leq$ and read `\emph{is included by}'.  The simplest example is the ordering of the integers according to the usual meaning of the symbol $\leq$ `\emph{is less than or equal to}'.  This results in a totally ordered structure called a \emph{chain} (Fig. \ref{fig:posets}A).  To illustrate the hierarchy, we simply draw element $B$ above element $A$ if $A \leq B$ and connect them with a line if there does not exist an element $X$ in the set such that $A \leq X \leq B$.

In some cases, elements of the set are incomparable to one another, as in the popular example of comparing apples and oranges.  A set of incomparable elements is called \emph{antichain}.  I illustrate this in Figure \ref{fig:posets}B with a set of card suits where the elements are placed side-by-side to indicate that no element includes any other.

More interesting examples involve both inclusion and incomparability, which is why we refer to these structures in general as \emph{partially ordered sets}, or \emph{posets} for short.  Figure \ref{fig:posets}C illustrates the poset that results from partitioning three objects.  One could consider all three objects together $a b c$, or each separately $a | b | c$.  These objects can also be partitioned in three ways: $a | b c$, $b | a c$ or $c | a b$.  Any two partitions from this set can be compared according to a relation that decides whether one partition includes another.  For example, the partition $a b c$ includes the partition $a | b | c$ since it can be obtained by simply sub-dividing $abc$ into three separate cells.  However, the partitions $c | a b$ and $a | b c$ are incomparable since, for example, there is no way to sub-divide the partition $c | a b$ to obtain the partition $a | b c$.

\begin{figure}[t]
  \label{fig:lattice}
  \includegraphics[height=.12\textheight]{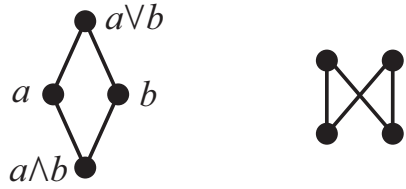}
  \caption{The poset on the left is a simple lattice, which illustrates the join $\join$ and the meet $\meet$.  The poset on the right is not a lattice since the pair of elements on the bottom do not have a unique least upper bound. Similarly, the pair of elements at the top do not have a unique greatest lower bound.}
\end{figure}

Given a set of elements in a poset, their \emph{upper bound} is the set of elements that contain each of the elements of the set.  For example, the upper bound of the partition $c | a b$ in Fig. \ref{fig:posets}C is the set $\{a b c\}$.  Given a pair of elements $x$ and $y$, the least element of their upper bound is called the \emph{join}, which is denoted $x \vee y$.  The \emph{lower bound} of a set of elements is defined dually by considering all the elements included by each of the elements of the set.  Given a pair of elements $x$ and $y$, the greatest element of their lower bound is called the \emph{meet}, which is denoted $x \wedge y$.  A \emph{lattice} is a partially ordered set where each pair of elements has a unique meet and a unique join (Fig. \ref{fig:lattice}).  Graphically, the join can be found by starting at both elements and following the lines upward until they first intersect.  The meet is found similarly by moving downward.  There often exist elements that are not formed from the join of any pair of elements.  These elements are called \emph{join-irreducible elements}.  \emph{Meet-irreducible elements} are defined similarly. For example, the partitions $a | b c$, $b | a c$ or $c | a b$ cannot be formed by joining any other pair of partitions and therefore are join-irreducible.  In this case, these elements are also meet-irreducible.

We can choose to view the join and meet as algebraic operations that take any two lattice elements to a unique third lattice element.  From this perspective, the lattice is an algebra.  This results in both a structural and operational perspective which are related by a set of equations called \emph{consistency relations}
\begin{equation}
\label{eq:lattice-algebra} x \leq y \qquad\Longleftrightarrow\qquad
   \begin{array}{rl}
       x \vee y = y \\
       x \wedge y = x \end{array}
\end{equation}

In short, a lattice is an algebra.  Where an algebra considers a set of elements along with a set of operations that takes one or more elements to another element, the lattice considers a set of elements along with a binary ordering relation that sets up a hierarchy among the elements.  The algebraic perspective is operational, whereas the lattice perspective is structural.  Both the operational and structural relationships among elements are useful.

Given a specific lattice, we find that the consistency relations result in a specific algebraic identity.
For example, the integers ordered by the usual `\emph{less than or equal to}' leads to
\begin{equation}
\label{eq:leq} x \leq y \qquad\Longleftrightarrow\qquad
   \begin{array}{rl}
       \max(x, y) = y \\
       \min(x, y) = x \end{array}
\end{equation}
whereas the positive integers ordered by `\emph{divides}' leads to
\begin{equation}
\label{eq:divides} y \mid x \qquad\Longleftrightarrow\qquad
   \begin{array}{rl}
       \mathrm{lcm}(x, y) = y \\
       \gcd(x, y) = x \end{array}
\end{equation}
Sets ordered by the usual `\emph{is a subset of}' leads to
\begin{equation}
\label{eq:subseteq} x \subseteq y \qquad\Longleftrightarrow\qquad
   \begin{array}{rl}
       x \cup y = y \\
       x \cap y = x \end{array}
\end{equation}
Such examples highlight the generality of the order-theoretic approach.

\section{Quantification}
There are many ways to quantify a poset.  Here I will describe some of the ways that we have been exploring \cite{Knuth:laws, Knuth:measuring, Knuth+Bahreyni:SR}: valuations, bi-valuations, and projections.  However, I will leave a more general discussion of the pair formalism of quantum mechanics and the origin of the complex sum and product rules as described in \cite{GKS:PRA} to a future work.  It is important to keep in mind that the quantification techniques I will cover does not comprise an exhaustive list, as we are only beginning to explore the possibilities.

We begin by considering the quantification of lattices.  We will see that this is equivalent to extending an algebra to a calculus by defining functions that take lattice elements to real numbers.  Such functions enable one to \emph{quantify} the relationships between the lattice elements.  This leads to probability theory on the lattice of logical statements and information theory on the partition sublattice of questions \cite{Knuth:measuring}.

\subsection{Valuations and Bi-valuations}
A valuation $v$ is a function that takes a single lattice element $x \in L$ to a real number $v(x)$ in a way that respects the partial order, so that $v(x) \leq v(y)$ iff $x \leq y$.  This means that the lattice structure imposes constraints on the valuation assignments, which can be expressed as a set of constraint equations.

The valuation assigned to element $x$ can be defined with respect to a second lattice element $y$ called the \emph{context}.  The result is a function called a bi-valuation $w(x \mid y) = v_y(x)$, which takes two lattice elements $x$ and $y$ to a real number.  Here a solidus is used as an argument separator so that one reads $w(x \mid y)$ as the degree to which $y$ includes $x$.

In the following sections, I consider three operations than can be performed on lattices, each of which obeys associativity.  The symmetries exhibited by associativity impose strong constraints on quantification, namely additivity.  This, in turn, constrains valuation and bi-valuation assignments.  The first two operations, the lattice join and the lattice product, are associated with the lattice structure and thus impose the same constraints on both the valuation and bi-valuation assignments; whereas the last symmetry, associativity of context, is specific to bi-valuations.

\begin{figure}
  \label{fig:sum}
  \includegraphics[height=.15\textheight]{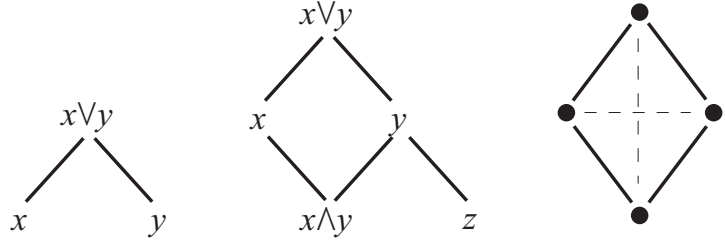}
  \caption{The poset on the left is used to establish the additive nature of the valuation.  The poset in the center is used to establish the sum rule for the lattice in general.  The cartoon on the right illustrates the symmetry of the sum rule.  The sum of the valuations of the elements at the top and bottom of the diamond equals the sum of the valuations of the elements on the right and left sides.  These dashed lines conveniently form a plus sign reminding us of the sum rule.}
\end{figure}

\subsubsection{The Lattice Join}
I now show that associativity of the lattice join forces valuations to be additive.  I begin by considering a very special case depicted in Fig. \ref{fig:sum} (left) of two elements $x$ and $y$ with join $x \join y$ and a null meet $x \meet y = \bot$ (not shown).  The value assigned to the join $x \join y$, written $u(x \join y)$, must be a function of the values assigned to both $x$ and $y$, $u(x)$ and $u(y)$, since if there did not exist any functional relationship, then the valuation could not possibly reflect the underlying lattice structure.  This functional relationship can be written in terms of an unknown binary operator $\oplus$
\begin{equation}
u(x \join y) = u(x) \oplus u(y).
\end{equation}
Now consider another case where we have three elements $x$, $y$, and $z$, such that their meets are again disjoint.  The least upper bound of these three elements can be written in at least two different ways: $x \join (y \join z)$ and $(x \join y) \join z$.  Consequently, the value assigned to this join can also be written in two different ways
\begin{equation}
u(x) \oplus \big(u(y) \oplus u(z)\big) = \big(u(x) \oplus u(y)\big) \oplus u(z).
\end{equation}
This functional equation for the operator $\oplus$ has a general solution given by Aczel \cite{Aczel:FunctEqns}
\begin{equation}
f(u(x \join y)) = f(u(x)) + f(u(y)),
\end{equation}
where $f$ is an arbitrary invertible function.
We take advantage of this freedom to choose a valuation $v(x) = f(u(x))$ that simplifies this constraint
\begin{equation} \label{eq:simple-sum}
v(x \join y) = v(x) + v(y).
\end{equation}
By letting $x = \bot$, equation (\ref{eq:simple-sum}) implies that $v(\bot) = 0$.

We now seek a solution for the general case.  Consider the lattice in Figure \ref{fig:sum} (center) and note that the elements $x \meet y$ and $z$ have a null meet, as do the elements $x$ and $z$.  Applying (\ref{eq:simple-sum}) to these two cases, we get
\begin{eqnarray}
v(y) & = & v(x \meet y) + v(z)\\
v(x \join y) & = & v(x) + v(z)
\end{eqnarray}
Simple substitution results in the general constraint equation known as the \emph{sum rule}
\begin{equation}
v(x \join y) = v(x) + v(y) - v(x \meet y).
\end{equation}
In general for bi-valuations we have
\begin{equation}
w(x \join y \mid t) = w(x \mid t) + w(y \mid t) - w(x \meet y \mid t).
\end{equation}
for any context $t$. Note that the sum rule is not focused solely on joins since it is symmetric with respect to interchange of joins and meets.  That is, this result simultaneously respects associativity of the lattice join and the lattice meet.

We have \emph{derived} that associativity constrains us to additive valuations---there is no other option.  The cartoon at the right of Fig. \ref{fig:sum} illustrates the symmetry of the sum rule.  The sum of the valuations of the elements at the top and bottom of the diamond equals the sum of the valuations of the elements on the right and left sides
\begin{equation} \label{eq:sum-rule}
v(x \join y) + v(x \meet y) = v(x) + v(y).
\end{equation}

\subsubsection{The Lattice Product}
One can combine two lattices via the lattice product where elements themselves are combined in as in a Cartesian product.  That is, the product of a lattice $X$ with a lattice $Y$ will result in a lattice $X \times Y$ with elements of the form $(x, y)$, where $x \in X$ and $y \in Y$.  The lattice product is associative, so that for three lattices $X$, $Y$, and $Z$, we have
\begin{equation}
(X \times Y) \times Z = X \times (Y \times Z)
\end{equation}
with elements of the form $(x,y,z)$.

The valuation assigned to an element $(x, y)$ clearly must be a function of the valuations assigned to $x$ and $y$ in their respective original lattices.  Again, associativity will require that they are combined in an additive fashion
\begin{equation}
g(u((x, y))) = g(u(x)) + g(u(y)),
\end{equation}
where $g$ is an arbitrary function.

In some cases, such as in probability theory, we expect associativity of the lattice product to hold simultaneously with associativity of the lattice join within a given lattice.  Given the linearity of the constraint imposed by associativity of lattice join (\ref{eq:sum-rule}), the only remaining freedom is that of rescaling.  This means that any further constraints must have a multiplicative form.  The result is that the valuation assigned to an element formed by a lattice product is given by
\begin{equation}
v((x, y)) = v(x) v(y),
\end{equation}
which is a \emph{product rule} applicable to combining lattices.

\subsubsection{The Chain Rule}
We now focus on bi-valuations and explore changes in context. Changes in context are again associative, which again results in an additive constraint.

\begin{figure}
  \label{fig:prod}
  \includegraphics[height=.35\textheight]{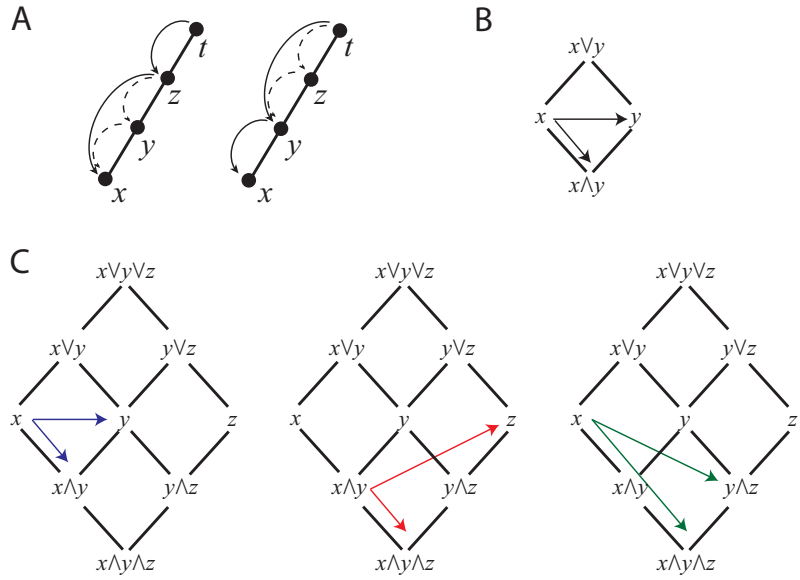}
  \caption{(A) Associativity of context is used to derive the chain rule.  (B) The diamond illustrates that the degree to which $x$ includes $x \meet y$ equals the degree to which $x$ includes $y$, $w(y\mid x) = w(x \land y\mid x)$.  (C) The lemma in panel B is used repeatedly to transform the chain rule into the usual product rule.}
\end{figure}

We begin with the special case of a chain and consider four ordered elements $x \le y \le z \le t$.  The relationship $x \le z$ can be divided into two relations, $x \le y$ and $y \le z$.  By considering $z$ to be the context, this sub-division implies that the context can be considered in parts.  Thus the bi-valuation we assign to $x$ with respect to context $z$, $w(x \mid z)$, must be related to both the bi-valuation assigned to $x$ with respect to context $y$, $w(x \mid y)$, and the bi-valuation assigned to $y$ with respect to context $z$, $w(y \mid z)$.  That is, there exists a binary operator $\odot$ that relates the bi-valuations assigned to the two steps to the bi-valuation assigned to the one step
\begin{equation}
    w(x\mid z) = w(x\mid y) \odot w(y\mid z)\,.
\end{equation}
Extending this to three steps (Fig. \ref{fig:prod}A) and considering the bi-valuation $w(x\mid t)$ relating $x$ and $t$, via intermediate contexts $y$ and $z$, we obtain another associative relationship
\begin{equation}
    \big( w(x\mid y) \odot w(y\mid z) \big) \odot w(z \mid t) = w(x\mid y) \odot \big( w(y\mid z) \odot w(z \mid t) \big)
\end{equation}
Using the associativity theorem again results in a constraint equation for non-negative bi-valuations involving changes in context \cite{Skilling:me08}.  We call this the \emph{chain rule}
\begin{equation}
w(x \mid z) = w(x \mid y) w(y \mid z)\,.
\end{equation}

This result can be extended by considering the following lemma.  The sum rule applied to the diamond in Fig. \ref{fig:prod}B defined by $x$, $y$, $x \lor y$, and $x \land y$ with context $x$ gives
\begin{equation}
w(x\mid x) + w(y\mid x) = w(x \lor y\mid x) + w(x \land y\mid x).
\end{equation}
Since $x \leq x$ and $x \leq x \lor y$, we have $w(x\mid x) = w(x \lor y\mid x) = 1$, reducing the sum rule to
\begin{equation}\label{eq:prodstep1}
w(y\mid x) = w(x \land y\mid x).
\end{equation}
This relationship, illustrated by the equivalence of the arrows in Fig. \ref{fig:prod}B, will used several times in the derivation that follows.

We now consider the more general lattice in Fig. \ref{fig:prod}C and
focus on the chain along the lower left side. Using the chain rule, we decompose the bi-valuation $w(x \meet y \meet z \mid x)$ with context $x$ into two parts by introducing the intermediate context $x \meet y$
\begin{equation}\label{eq:prodchain}
w(x \meet y \meet z\mid x) = w(x \meet y \meet z \mid x \meet y)\,w(x \meet y \mid x).
\end{equation}
We apply the lemma to the diamond defined by $x \meet y \meet z$, $x \meet y$, $y \meet z$, $z$ (Fig. \ref{fig:prod}C, center) to obtain
\begin{equation}\label{eq:prodstep2}
w(x \meet y \meet z \mid x \meet y) = w(z \mid x \meet y).
\end{equation}
Similarly, the diamond defined by $x$, $x \meet y$, $y \meet z$, and $x \meet y \meet z$ (Fig. \ref{fig:prod}C, right) results in
\begin{equation}\label{eq:prodstep3}
w(x \meet y \meet z \mid x) = w(y \meet z \mid x).
\end{equation}
Substituting (\ref{eq:prodstep1}),(\ref{eq:prodstep2}), and (\ref{eq:prodstep3}) into (\ref{eq:prodchain}) results in the \emph{product rule} for context change.
\begin{equation}
w(y \meet z \mid x) = w(z \mid x \meet y)\,w(y \mid x).
\end{equation}

\subsubsection{The Valuation Calculus}
We have derived that associativity of the lattice join results in the sum rule
\begin{equation}
    v(x \join y) + v(x \meet y) = v(x) + v(y)\,,
\end{equation}
which is a central axiom of measure theory.  Associativity of the lattice product imposes an additional constraint, which results in a product rule
\begin{equation}
    v((x, y)) = v(x)v(y)\,.
\end{equation}

Extending the concept of valuation to that of a context-dependent bi-valuation, we obtain a sum rule
\begin{equation}
    w(x \join y \mid t) + w(x \meet y \mid t) = w(x \mid t) + w(y \mid t)\,,
\end{equation}
a product rule for combining spaces
\begin{equation}
    w((x, y) \mid (t_x, t_y)) = w(x \mid t_x)w(y \mid t_y)\,,
\end{equation}
and a product rule for context change
\begin{equation}
w(y \meet z \mid x) = w(z \mid x \meet y)\,w(y \mid x)\,.
\end{equation}
The valuation calculus differs from traditional measure theory in two important ways.  First, additivity is not postulated, but rather is derived from associativity.  Second, the valuation calculus generalizes measure theory by introducing the concept of context, which is quantified using bi-valuations and manipulated using the product rule.  These rules are constraint equations ensuring that the assigned valuations respect the order-theoretic properties of the lattice.

\begin{figure}
  \label{fig:projection}
  \includegraphics[height=.2\textheight]{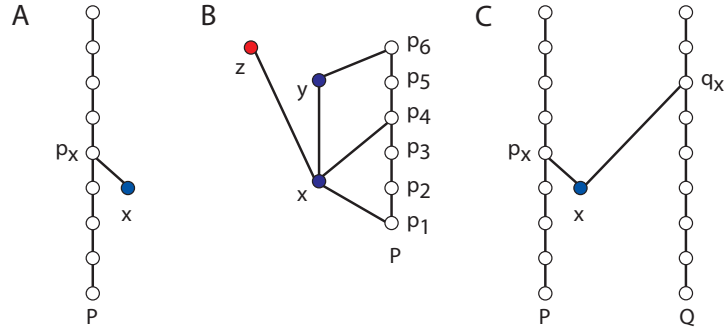}
  \caption{(A) The projection of an event $x$ onto a chain is the least event on the chain that includes $x$. (B) In this poset, elements $x$ and $y$ are quantifiable by the chain $P$, whereas element $z$ is not.  The number of distinct quantifiable classes of elements is given by the number of top elements of the poset.  (C) Multiple chains can be used to quantify poset elements.  Here the element $x$ is quantified by the numeric pair $(p_x, q_x)$.}
\end{figure}

\subsection{Projections}
The previous sections describe the consistent quantification of lattices, which is made possible by the fact that lattices possess extra structure that allows one to define a unique join and meet of each pair of elements thus making it an algebra.  It is precisely this extra structure that constrains any proposed quantification scheme via the sum and product rules.  However, such constraints do not apply to posets in general since they lack this extra structure possessed by lattices.

Consistent quantification of a poset can proceed by artificially imposing additional lattice-like structure.  One way to do this is to select a \emph{distinguished a set of elements} in the poset that form a lattice, and attempt to relate the remaining elements in the poset to the elements of this \emph{distinguished set}. We have recently demonstrated this quantification technique by selecting one or more chains as the distinguished set (or sets) and \emph{projecting} poset elements onto the chains \cite{Knuth+Bahreyni:SR}.  In general, it may not be possible to quantify all poset elements in this way, but here we show that one can certainly quantify a subset of the elements. Surprisingly, this proposed quantification scheme results in the Minkowski metric and Lorentz transformations \cite{Knuth+Bahreyni:SR}.

\subsubsection{Coordinates}
First we consider quantification using a single chain.  We select a chain $P$ to be used for quantification and label its elements with $i$.  In a finite poset, such a chain is described by $p_1 \leq p_2 \leq \ldots \leq p_i \leq \ldots p_N$. In an infinite poset where the chain is countably infinite the label $i$ can be any integer and the chain is described by $\dots \leq p_{i-1} \leq p_i \leq p_{i+1} \leq \ldots$.  If the chain is uncountably infinite, a real number index can be used.

An element $x$ can be projected onto a chain $P$ if there exists an element $p \in P$ such that $x \leq p$.  If this is the case, then the \emph{projection} of $x$ onto the chain $P$ is given by the least element $p_x$ on the chain $P$ such that $x \leq p_x$.  If one considers the sub-poset consisting only of the element $x$ and the elements comprising the chain $P$, then in this sub-poset $p_x$ covers $x$, $p_x \succ x$ (Fig. \ref{fig:projection}A).  If the projection exists, we say that $x$ is \emph{quantifiable} with respect to $P$, and assign to the element $x$ the numeric label assigned to the element $p_x \in P$.  Note that, in general, not all elements of a poset are quantifiable with respect to a given chain.  Any chain potentially divides the poset into two classes: elements quantifiable with respect to the chain and elements not quantifiable with respect to the chain (Fig. \ref{fig:projection}B).  Thus, one can only be assured to quantify some subset of the poset.

One can project to N different chains and use the corresponding numeric labels to \emph{coordinatize} the poset elements that are quantifiable with respect to each of the selected chains with numbers taken as a Cartesian product (Fig. \ref{fig:projection}C).

\begin{figure}
  \label{fig:intervals}
  \includegraphics[height=.2\textheight]{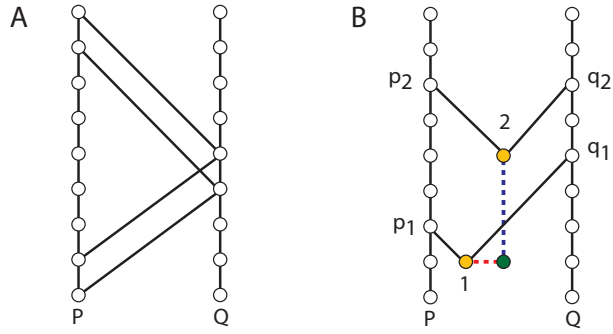}
  \caption{(A) Chains can be synchronized by selecting quantifying elements such that successive elements on one chain project to successive elements on the other, and vice versa. (B) This illustrates a method to quantify an interval between two poset elements as well as its decomposition into a symmetric (chain-like) part and an anti-symmetric (antichain-like) part.  Chain-like relationships are analogous to time-like relationships; whereas antichain-like relationships are analogous to space-like relationships.}
\end{figure}

\subsubsection{Intervals}
The \emph{interval} between two poset elements can be quantified using two chains.  These chains must be \emph{synchronized} so that successive events in one chain project to successive events in the other chain (Fig. \ref{fig:intervals}A).  Figure \ref{fig:intervals}B illustrates the quantification of an interval given by $(\Delta p, \Delta q)$ where $\Delta p = p_2 - p_1$ and $\Delta q = q_2 - q_1$.  This pair-wise quantification can be decomposed into the sum of a symmetric and an antisymmetric pair \cite{Knuth+Bahreyni:SR} given by
\begin{equation}
(\Delta p, \Delta q) = \Big(\frac{\Delta p + \Delta q}{2}, \frac{\Delta p + \Delta q}{2} \Big) + \Big(\frac{\Delta p - \Delta q}{2}, \frac{\Delta q - \Delta p}{2} \Big)
\end{equation}

The two integer labels can be used to obtain a single scalar.  This is done by taking the lattice product of the two chains, which, as we saw earlier, results in a valuation found by taking the product of the two original valuations, so that
\begin{equation}
\Delta s^2 = \Delta p \Delta q.
\end{equation}
By defining
\begin{eqnarray}
\Delta t & = & \frac{\Delta p + \Delta q}{2} \\
\Delta x & = & \frac{\Delta p - \Delta q}{2}
\end{eqnarray}
we can rewrite the pair as
\begin{equation}
(\Delta p, \Delta q) = (\Delta t, \Delta t) + (\Delta x, -\Delta x)
\end{equation}
and the scalar as
\begin{equation}
\Delta s^2 = \Delta t^2 - \Delta x^2.
\end{equation}
This is the Minkowski metric, familiar from special relativity, and here it arises from a simple method for quantifying a poset \cite{Knuth+Bahreyni:SR}.  This is not a coincidence.  Our recent paper demonstrates that the scalar interval $\Delta s^2$ is invariant when computed with respect to any synchronized pair of chains.  In addition, the parameters $\Delta t$ and $\Delta x$ are shown to transform according to the Lorentz transformations of time and space.

It should be noted that such a consistent decomposition of an interval is not always possible given more than two synchronized chains \cite{Knuth+Bahreyni:SR}, and that this is related to the multi-dimensionality of space.

\section{Applications}
It is not possible in this tutorial to cover the applications derived using this methodology in requisite detail.  For this reason, I will simply outline the basic applications and point to appropriate references.  Since these quantification techniques are applicable to a wide array of posets and lattices, we can expect that they will be relevant to numerous applications.  At this point, we have five examples where we have derived a theory from first principles based on quantifying posets and lattices.

The most general of these applications, measure theory, has been discussed here as the derivation of the valuation calculus and the related bi-valuations.  The valuation calculus both encompasses and extends traditional measure theory.  Additivity of measures, which is an axiom of measure theory is derived here as a consequence of associativity.  Furthermore, the valuation calculus generalizes measure theory by introducing the concept of context.  A valuation with respect to a context is quantified using bi-valuations and manipulated using the product rule.  Earlier works discussing these results can be found here \cite{Knuth:laws, Knuth:measuring}.

The second example, which was the original inspiration for this work is the derivation of probability theory \cite{Knuth:laws, Knuth:lattices+probability, Knuth:me08, Knuth:measuring}.  By founding probability theory as a quantification of implication among logical statements, we obtain a theory that encompasses and generalizes both the Cox and Kolmogorov formulations.  By introducing probability as a bi-valuation defined on a lattice of statements we can quantify the degree to which one statement implies another.  Rather than deriving probability theory from a set of desiderata derived from Cox's particular notion of plausibility, the properties of the lattice of statements form the basis of the theory.  Furthermore, the \emph{meaning} of the derived measure is inherited from the ordering relation, which in this case is implication.  The fact that these lattices are derived from sets means that this work encompasses Kolmogorov's formulation of probability theory as a measure on sets.  However, mathematically this theory improves on Kolmogorov's foundation by not only \emph{deriving}, rather than assuming, additivity of the measure, but also by introducing the concept of context and endowing the measure with meaning.

The third example involves the derivation of information theory as a valuation on the partition subspace of questions.  The space of questions is generated from the space of statements by virtue of Birkhoff's Representation Theorem \cite{Knuth:Questions}.  The result is the free distributive lattice of questions, which by virtue of its being a lattice imposes a sum rule and a product rule.  By postulating that the relevance of a question is a function of the probabilities that answer it, we couple the probability measure on the statement space with the relevance measure on the question space.  Due to a conflict of constraints, to be discussed in more detail in a future work, one can show that an objective non-trivial measure can be defined only on the subspace of questions that are isomorphic to partitions.  The result is that the most basic relevance measures are quantified by the Shannon entropy of the set of assertions that potentially answer the question.  The sum rule, when relating partitions, results in a relationship between mutual information and joint entropy
\begin{equation}
I(A;B) = H(A) + H(B) - H(A,B).
\end{equation}
The result is not only a novel derivation of information theory, but a natural extension of the theory to include the relevance of a question quantified with respect to a given context \cite{Knuth:Questions, Knuth:WCCI06, Knuth:me08}.

Deriving mathematical theories is one thing, but deriving physical theories is an another thing altogether.  The first such example is a derivation of the complex sum and product rules of the Feynman formulation of quantum mechanics \cite{GKS:me09, GKS:PRA}.  This was achieved by considering a pair-wise valuation on the space of sequences of measurements.  The logic of the process of measuring served to generate the algebra, which implicitly defines a poset of measurement sequences.  By combining measurements in two ways: parallel and serial, which correspond to the lattice join and the lattice product, and mapping the pair-wise valuation to a scalar-valued probability, we obtain the complex sum and product rule along with the Born rule, which maps our pair-wise valuation to a scalar-valued probability \cite{GKS:me09, GKS:PRA}.

The most recent application has been a derivation of special relativity as a quantification of a poset of causally related events \cite{Knuth+Bahreyni:SR}.  As discussed above, this is achieved by distinguishing two chains of elements (events) as observers and projecting events onto the observer chains.  The result is that intervals are quantified by a pair of numbers and that this pair maps to a unique scalar, which gives rise to the Minkowski metric.  What is strange is that in this picture space and time emerge as nothing more than a convenient decomposition, which along with other results, strongly suggests that they are not fundamental.

\section{Conclusion}
In his derivation of probability theory Cox provided the first example of generalizing an algebra to a calculus \cite{Cox:1946}.  That such an activity is generally possible or even useful is not obvious until one begins to notice the great many similarities between a variety of mathematical theories and physical laws, such as the various incarnations of the sum rule or the fact that quantum mechanics looks like a complex version of probability theory.  As Jaynes recognized, it is not a matter of simple analogy, but rather something far more subtle.  The theories are similar because \emph{the ideas that lead to the theories are similar}.  These ideas are based on the quantification of order.

In this tutorial, I have shown how a variety of rules involving quantification arise as constraint equations to ensure that any quantification does not violate the underlying order.  What is more striking is that this entire procedure is based on the quantification of order underlying our descriptions of physical reality---not necessarily physical reality itself.  The consequence is that the physical laws we obtain are constraints on quantification imposed by our descriptions.  This is where we arrive at Information Physics.

At the heart of this new methodology lies the valuation calculus which is applicable to any lattice.  Associativity of the lattice join (or meet) gives rise to the sum rule.  Associativity of the lattice product results in a product rule, which dictates how valuations are to be combined when taking lattice products.  Associativity of changes of context result in a product rule for bi-valuations that dictates how valuations should be manipulated when changing context.  The techniques based on projections are based on distinguishing a sub-lattice that can be used to employ valuations to quantify a poset in general.

Most exciting is the range of theories that have been successfully derived using this foundation: measure theory, probability theory, information theory, quantum mechanics, and special relativity.  These results provide strong support for the claim that Information Physics, which relies on information about our descriptions of reality to derive physical laws, is a potentially useful general approach.  With these positive examples as guideposts, we now aim to use these techniques to quantify new problems and derive new physical laws.

%%%%%%%%%%%%%%%%%%%%%%%%%%%%%%%%%%%%%%%%%%%%%%%%
%% BACKMATTER
%%%%%%%%%%%%%%%%%%%%%%%%%%%%%%%%%%%%%%%%%%%%%%%%

\begin{theacknowledgments}
I would like to thank Janos Acz\'{e}l, Newshaw Bahreyni, Ariel Caticha, Julian Center, Seth Chaiken, Keith Earle, Adom Giffin, Philip Goyal, Steve Gull, Jeffrey Jewell, Vassilis Kaburlasos, Nabin Malakar, Carlos Rodr\'{i}guez, and John Skilling for inspiring discussions, invaluable remarks and comments, and much encouragement.  This work was supported in part by the College of Arts and Sciences and the College of Computing and Information of the University at Albany (SUNY).
\end{theacknowledgments}

%%%%%%%%%%%%%%%%%%%%%%%%%%%%%%%%%%%%%%%%%%%%%%%%
%% The bibliography can be prepared using the BibTeX program or
%% manually.
%%
%% The code below assumes that BibTeX is used.  If the bibliography is
%% produced without BibTeX comment out the following lines and see the
%% aipguide.pdf for further information.
%%
%% For your convenience a manually coded example is appended
%% after the \end{document}
%%%%%%%%%%%%%%%%%%%%%%%%%%%%%%%%%%%%%%%%%%%%%%%%

%%%%%%%%%%%%%%%%%%%%%%%%%%%%%%%%%%%%%%%%%%%%%%%%
%% You may have to change the BibTeX style below, depending on your
%% setup or preferences.
%%
%%
%% For The AIP proceedings layouts use either
%%%%%%%%%%%%%%%%%%%%%%%%%%%%%%%%%%%%%%%%%%%%

\bibliographystyle{aipproc}   % if natbib is available
%\bibliographystyle{aipprocl} % if natbib is missing

%%%%%%%%%%%%%%%%%%%%%%%%%%%%%%%%%%%%%%%%%%%
%% You probably want to use your own bibtex database here
%%%%%%%%%%%%%%%%%%%%%%%%%%%%%%%%%%%%%%%%%%%
\bibliography{knuth}

%%%%%%%%%%%%%%%%%%%%%%%%%%%%%%%%%%%%%%%%%%%
%% Just a reminder that you may have to run bibtex
%% All of it up to \end{document} can be removed
%% if you don't like the warning.
%%%%%%%%%%%%%%%%%%%%%%%%%%%%%%%%%%%%%%%%%%%
\IfFileExists{\jobname.bbl}{}
 {\typeout{}
  \typeout{******************************************}
  \typeout{** Please run "bibtex \jobname" to optain}
  \typeout{** the bibliography and then re-run LaTeX}
  \typeout{** twice to fix the references!}
  \typeout{******************************************}
  \typeout{}
 }

\end{document}